\documentclass[preprint,5p]{elsarticle}

\usepackage{url}
\usepackage[draft
              ]{hyperref}
\usepackage{amssymb,amsmath}
\usepackage{color}
\usepackage[normalem]{ulem}

\biboptions{sort&compress}

\newcommand{\gev}{\ \mathrm{GeV}}
\newcommand{\tev}{\ \mathrm{TeV}}

\newcommand{\pb}{\ \mathrm{pb}}
\newcommand{\neut}[1]{\tilde{\chi}^0_{#1}}
\newcommand{\cha}[1]{\tilde{\chi}^\pm_{#1}}
\newcommand{\sstop}{\tilde{t}_1}
\newcommand{\mstop}[1]{m_{\tilde{t}_{#1}}}
\newcommand{\mneu}[1]{m_{\tilde{\chi}^0_{#1}}}
\newcommand{\mcha}[1]{m_{\tilde{\chi}^\pm_{#1}}}

\begin{document}

\begin{frontmatter}

\title{Refining light stop exclusion limits with $W^+W^-$ cross sections}
\tnotetext[preprints]{IFT-UAM/CSIC-15-052}

\author[kr]{Krzysztof Rolbiecki}
\ead{rolbiecki.krzysztof@csic.es}

\author[jt]{Jamie Tattersall}
\ead{j.tattersall@thphys.uni-heidelberg.de}

\address[kr]{Instituto de F\'{\i}sica Te\'{o}rica UAM/CSIC, Madrid, 
  Spain}
\address[jt]{Institut f\"ur Theoretische Physik, University of Heidelberg, Heidelberg, Germany}

\begin{abstract}
 
If light supersymmetric top (stop) quarks are produced at the LHC and decay via on- or off-shell $W$-bosons they can be expected to 
contribute to a precision $W^+W^-$ cross section measurement. Using the latest results of the CMS experiment, we 
revisit constraints on the stop quark production and find that this measurement can exclude portions of the parameter 
space not probed by dedicated searches. In particular we can exclude light top squarks 
up to 230~GeV along the line separating three- and four-body decays, $\tilde{t}_1 \to \tilde{\chi}_1^0 W^{(*)} b$. We 
also study the exclusion limits in case when the branching ratio 
for these decays is reduced and show significant improvement over previously existing limits.
 
\end{abstract}


\end{frontmatter}

\section{Introduction\label{sec:intro}}

Searches for stops --- the supersymmetric (SUSY) partners of top quarks ---  have received significant attention from both 
ATLAS~\cite{Aad:2013ija,Aad:2014kra,Aad:2014bva,Aad:2014qaa,Aad:2014nra,Aad:2014mfk} and 
CMS~\cite{CMS-PAS-SUS-13-009,CMS-PAS-SUS-13-015,Chatrchyan:2013xna,Khachatryan:2015pwa,Khachatryan:2015wza,CMS-PAS-SUS-14-021}. While 
limits obtained after Run 1 of the LHC at $\sqrt{s} = 8\tev$ can go, depending on the decay modes studied, 
up to 800~GeV, there are still parts of parameter space where relatively light 
stops are allowed, see e.g.\ the summary plots by ATLAS~\cite{atlas_combined} and CMS~\cite{cms_combined}. 

The main motivation for light stops is the so-called natural supersymmetry~\cite{Papucci:2011wy} paradigm which demands
that the particles must be close in mass to the ordinary top quark. Unfortunately however, this region of parameter space 
is particularly difficult to explore due to the background of top quark production. In particular
if the stop quark decays via a top quark that is almost on-shell ($\tilde{t}_1 \to \tilde{\chi}^0_1 t^{(*)}$), no 
exclusion limit is currently present. Another difficult region of the parameters space can be identified at 
the border between three- and four-body decays 
with a (nearly) on-shell $W$ boson ($\tilde{t}_1 \to \tilde{\chi}^0_1 W^{(*)} b$).

Several recent theoretical studies have attempted to fill these holes in the stop 
parameter space by using precise predictions and measurements of top quark cross 
section~\cite{Czakon:2014fka} (see however ref.~\cite{Eifert:2014kea} for a discussion of possible problems with this approach), specialized mono-jet searches~\cite{Ferretti:2015dea}, recasting other SUSY searches~\cite{Delgado:2012eu} or via angular 
correlations~\cite{Buckley:2014fqa}. A complementary idea is that certain corners of the parameter space might be constrained by 
looking for signals of stoponium production~\cite{Batell:2015zla}.

An alternative approach presented here is based on the observation that light stops decaying into certain final 
states can
contribute to the $W^+W^-$ cross section measurements~\cite{Curtin:2012nn,Rolbiecki:2013fia,Kim:2014eva,Curtin:2014zua,Dermisek:2015vra}. Until recently 
the ATLAS and CMS results were displaying a moderate excess over the 
standard model (SM) prediction~\cite{ATLAS:2012mec,Chatrchyan:2013oev,ATLAS-CONF-2014-033} 
but this was determined to be the result of neglected higher order corrections \cite{Gehrmann:2014fva,Jaiswal:2014yba,Meade:2014fca}.
In any case, the fact that the observed cross-section was greater than the predicted background meant 
that any derived constraints on stop production would have been weak. However, the recent 
CMS measurement~\cite{CMS-PAS-SMP-14-016} based on the full $\sqrt{s} = 8 \tev$ dataset, 
using the next-to-next-to-leading-order (NNLO) cross section 
prediction, $\sigma^{\mathrm{NNLO}}(pp\to W^+W^-) = 59.8^{+1.3}_{-1.1} \pb$~\cite{Gehrmann:2014fva}, and 
event re\-weighing~\cite{Meade:2014fca} turned out to be very well aligned with 
the SM: $\sigma^{\mathrm{exp}} = 60.1 \pm 4.8 \pb$. In this Letter, we recast the CMS analysis as a potential way to
constrain the production of light stops. 

We focus on three widely studied decay modes that are commonly present in SUSY models with light stops and improve the existing constraints. Assuming 
that only the light stop and the lightest supersymmetric particle (LSP, in our case the lightest 
neutralino, $\neut{1}$) have masses of order of the electroweak symmetry breaking (EWSB) scale we have:
\begin{align}
  &  \tilde{t}_1 \to \neut{1}\, t \,,  & &\mathrm{if}\ \ \mstop{1} \geq m_t + \mneu{1} \,,  \label{eq:2body}\\
  &  \tilde{t}_1 \to \neut{1}\, W\, b\,,  & &\mathrm{if}\ \ \mstop{1} \geq m_W + m_b + \mneu{1} \,, \label{eq:3body}\\
  &  \tilde{t}_1 \to  \neut{1}\, f\, f'\, b\,,  & &\mathrm{if}\ \ \mstop{1} < m_W + m_b + \mneu{1} \,.\label{eq:4body}
\end{align}
The three- and four-body decays might compete with loop-mediated two-body 
decay, $\tilde{t}_1 \to \neut{1}\,c$~\cite{Hikasa:1987db,Muhlleitner:2011ww}, but the branching ratios (BR) are 
highly model dependent here~\cite{Porod:1998yp,Boehm:1999tr,Grober:2014aha,Grober:2015fia}. Another possibility is given by:
\begin{align}
  &  \sstop \to \cha{1}\, b \to \neut{1}\, W^{(*)}\, b\,, \label{eq:2body_cha}
\end{align}
provided $\mcha{1} < \mstop{1}$. Depending on the parameter point under consideration, in particular 
the mass differences between stop and electroweakinos and their mixing character, the 
chargino mediated~\eqref{eq:2body_cha} and one of the direct decays~\eqref{eq:2body}, \eqref{eq:3body} or \eqref{eq:4body}, may 
be simultaneously present, see e.g.~\cite{Rolbiecki:2009hk}. This feature would have a significant impact on the expected exclusion limits.

\section{Simulation\label{sec:mc}}

Monte-Carlo stop samples with up to one additional jet 
were simulated using \texttt{MadGraph5\_aMC@NLO}~\cite{Alwall:2014hca} and matched 
to the \texttt{Pythia~6}~\cite{Sjostrand:2006za} parton shower. The cross sections were normalized to 
the next-to-leading-order (NLO) prediction using \texttt{NLLfast}~\cite{Beenakker:1997ut,Beenakker:2010nq}. 

The analysis of the simulated samples was performed using \texttt{CheckMATE}~\cite{Drees:2013wra,Kim:2015wza} and a 
dedicated implementation of the CMS $W^+W^-$ cross-section 
measurement. %, that was validated using the actual CMS numbers for $W^+W^-$ and $t\bar{t}$ production~\cite{CMS-PAS-SMP-14-016}. 
\texttt{CheckMATE} uses a specially tuned version of the \texttt{Delphes 3}
detector simulation \cite{deFavereau:2013fsa} and jets were clustered using  \texttt{FastJet}~\cite{Cacciari:2011ma} with 
the anti-$k_T$ algorithm \cite{Cacciari:2008gp}.  The analysis 
is performed for di-lepton final states with missing energy which for the signal process 
originates from neutrinos. In order to suppress a dominant SM background, $t\bar{t}$ production, events with 
$b$-jets and multiple final-state jets are vetoed. To better understand the $t\bar{t}$ background, CMS 
defines two signal regions (SR): 0-jet 
SR without jets with $p_T > 30 \gev$; 1-jet SR with exactly one jet with $p_T > 30 \gev$. These are further subdivided
based on whether the final state leptons have  different ($e\mu$) or same flavour ($ee$ or $\mu\mu$).
The expected and 
observed event numbers agree well within errors for all SR and therefore the analysis can serve as a constraint for models that 
contribute to the similar final state. 

Our implementation was validated using the event numbers 
provided by the CMS collaboration for $W^+W^-$ signal and $t\bar{t}$ background~\cite{CMS-PAS-SMP-14-016}. The samples used for validation were obtained using \texttt{MadGraph5\_aMC@NLO}~\cite{Alwall:2014hca} and hadronised using \texttt{Herwig++~2.7}~\cite{Bahr:2008pv,Bellm:2013lba}.
The parton distribution function (PDF) sets used were CTEQ6L~\cite{Nadolsky:2008zw} for leading order generation and 
CT10~\cite{Lai:2010vv} for NLO.  

 In order to apply limits to stop production we implement two different procedures. For the first we 
calculate the model independent confidence limits in a modified frequentist approach (CLs method \cite{0954-3899-28-10-313}) at 95\% for the two flavour inclusive SRs. We then define the 
stop model as excluded when it predicts a cross section in excess of any of these limits. The second method produces a more 
stringent model dependent CLs limit at 95\% by performing a combined fit to the four separate signal regions. We 
use \texttt{HistFitter}~\cite{Baak:2014wma} as an interface 
to \texttt{HistFactory}~\cite{Cranmer:2012sba}, \texttt{RooStats}~\cite{Moneta:2010pm}, \texttt{RooFit}~\cite{Verkerke:2003ir} and 
\texttt{ROOT}~\cite{Brun:1997pa,Antcheva:2011zz},
to allow all background sources to float independently whilst including the correct correlated systematics.

Many of the backgrounds to $W^+W^-$ production are determined with a data driven technique in the CMS 
analysis. Essentially control regions are defined for the various backgrounds that are only expected to contain a small
contribution from the $W^+W^-$ process under study. These are used to normalise the various backgrounds and Monte-Carlo is then
used to extrapolate to the $W^+W^-$ signal regions. One may therefore worry that stop production contributes in these control
regions, spoiling the normalisation constants.

Unfortunately CMS does not publish the control regions used so it is impossible for us to explore these effects. However, 
we note that it is possible that stop production contributes to the control region and thus increases
the value of the normalisation constant. In turn this could increase the 
background prediction in the $W^+W^-$ signal regions. Such an effect would actually strengthen the limit we derive on stop production
and one may worry about setting a spurious exclusion but  we believe that such a possibility does not exist in our 
study. Firstly, the predicted and measured $W^+W^-$ cross section are now in very good agreement suggesting that 
 any signal contamination can only be slight. Secondly, of most concern for this analysis is the $t\overline{t}$ control region. However,
in the parameter region of most interest for our study, $m_{\tilde{t}}\approx m_{b}+m_{W}+m_{\tilde{\chi}^0}$, this corresponds to 
the $b$-quarks being extremely soft. Hence it is very unlikely, that this final state will contribute  to a  $t\overline{t}$ measurement at all.

\section{Results\label{sec:results}}

\begin{figure*}[t]
\begin{center}\vspace*{-1.5cm}
\includegraphics[width=1.0\textwidth]{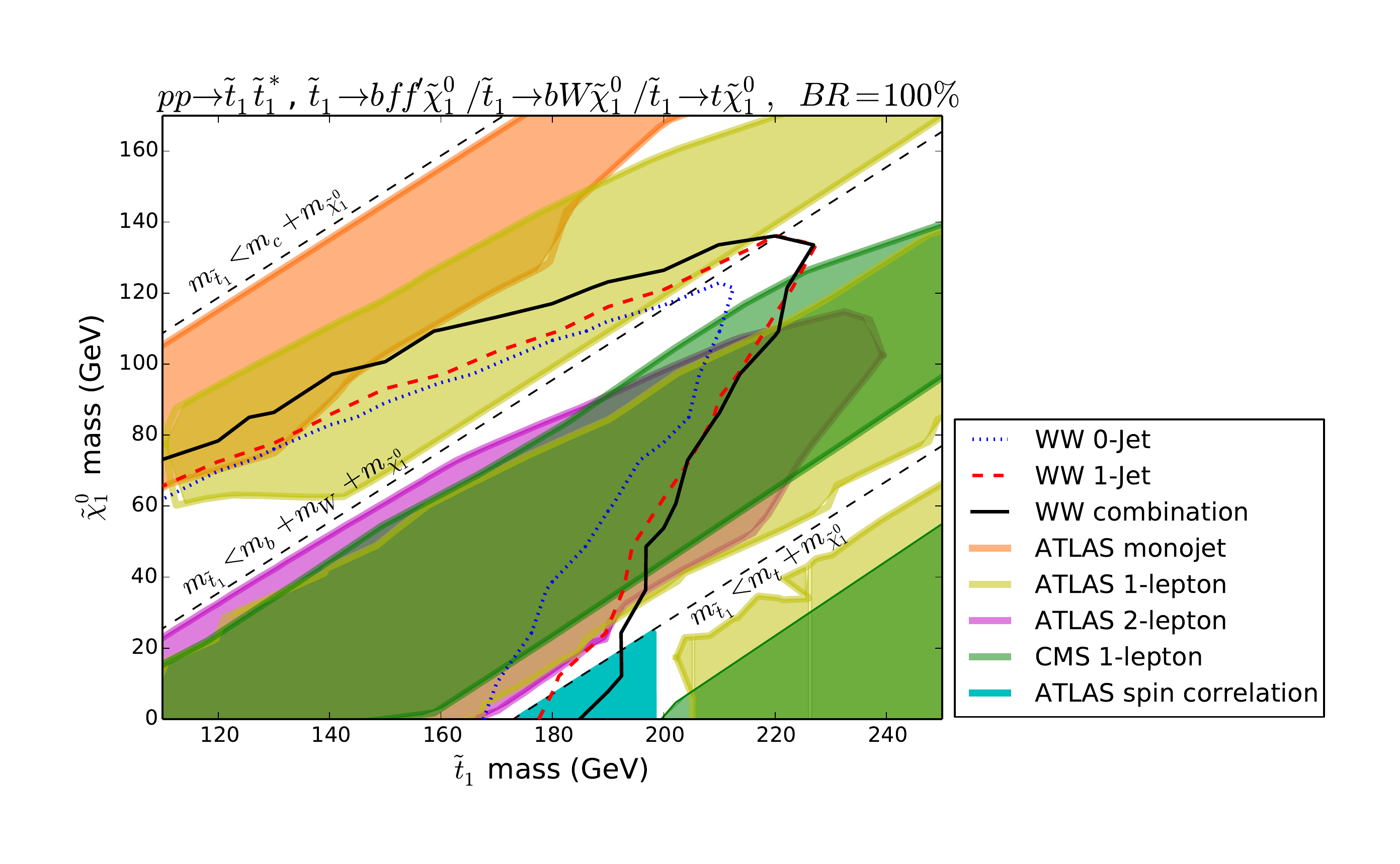}
\end{center}\vspace*{-1.5cm}
\caption{The exclusion limits for stop pair-production in $\mstop{1}$-$\mneu{1}$ plane assuming that only decay modes \eqref{eq:2body}-\eqref{eq:4body} are allowed, respectively. The dotted-blue line denotes the exclusion using 0-jet signal region and the red-dashed 1-jet signal region of Ref.~\cite{CMS-PAS-SMP-14-016}. The black solid-line is for the combined exclusion, as discussed in Sec.~\ref{sec:mc}. The experimental exclusions were extracted from the following studies: ATLAS monojet~\cite{Aad:2014nra}, ATLAS 1-lepton~\cite{Aad:2014kra}, ATLAS 2-lepton~\cite{Aad:2014qaa}, CMS 1-lepton~\cite{Chatrchyan:2013xna}, ATLAS spin correlation~\cite{Aad:2014mfk}.
 \label{fig:excl3body}}
\end{figure*}

\begin{figure*}[t]
\begin{center}\vspace*{-0.8cm}
\includegraphics[width=1.0\textwidth]{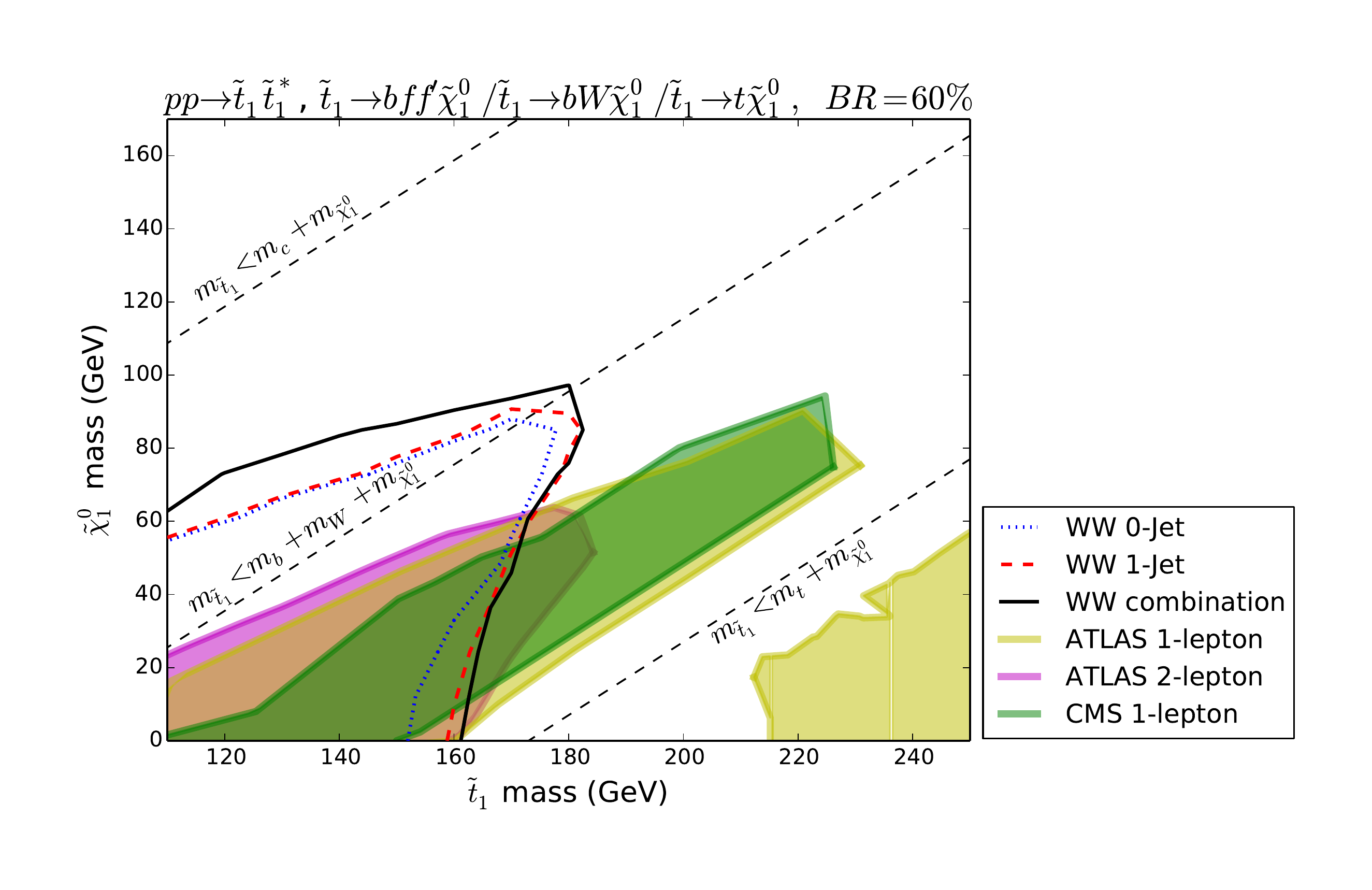}
\end{center}\vspace*{-1.3cm}
\caption{The exclusion limits for stop pair-production in $\mstop{1}$-$\mneu{1}$ plane assuming that the branching ratio for decay modes \eqref{eq:2body}-\eqref{eq:4body} is $60\%$. The dotted-blue line denotes the exclusion using 0-jet signal region and the red-dashed 1-jet signal region of Ref.~\cite{CMS-PAS-SMP-14-016}. The black solid-line is for the combined exclusion, as discussed in Sec.~\ref{sec:mc}. The experimental exclusions were extracted from the following studies: ATLAS monojet~\cite{Aad:2014nra}, ATLAS 1-lepton~\cite{Aad:2014kra}, ATLAS 2-lepton~\cite{Aad:2014qaa}, CMS 1-lepton~\cite{Chatrchyan:2013xna}, ATLAS spin correlation~\cite{Aad:2014mfk}.
 \label{fig:excl3body_06}}
\end{figure*}

The obtained exclusion limits for stop decaying via~\eqref{eq:2body}, \eqref{eq:3body} or 
\eqref{eq:4body} with 100\% branching ratio 
are shown in Fig.~\ref{fig:excl3body}. The best exclusion limit is 
obtained along the line $\mstop{1} \simeq m_W + m_b + \mneu{1}$ and is exactly where no
current LHC search sets a limit on these models. The $W^+W^-$ measurement 
allows us to constrain stop masses up to $\sim 220$~GeV for a LSP mass of $\sim 130$~GeV. We also note the additional exclusion for stop masses 
$\sim 170$--$190$~GeV for decay mode~\eqref{eq:3body} where an intermediate top quark is nearly on-shell that also extends the 
limits from dedicated stop searches.

The reason that the $W^+W^-$ cross section measurement is so sensitive along
this line is that the final states are most similar to the actual SM production of $W^+W^-$ pairs: 
the $W$ boson is (nearly) on-shell and the $b$-jet is rather soft, significantly 
reducing $b$-jet veto effectiveness. On the other hand, this region is problematic for 
dedicated stop searches due to its similarity to the SM background. Dedicated stop searches
attempt to place cuts that act as a discriminator between signal and background. However in regions where the signal 
has very similar features to the background this approach breaks down and consequently our 
approach is complementary to other searches.

It was shown in Ref.~\cite{Grober:2015fia} that the branching ratio for 
decays \eqref{eq:3body} and \eqref{eq:4body} can be substantially reduced in 
favour of loop-mediated flavour-changing two-body decay, $\tilde{t}_1 \to \neut{1}\,c$. Such a 
reduction can pose a significant challenge for dedicated stop searches as can be seen for 
example in Fig.~12 of Ref.~\cite{Chatrchyan:2013xna}. Therefore, we compare exclusion limits 
obtained in the current study with the results reported by collaborations, 
but now assume $BR(\tilde{t}_1 \to \tilde{\chi}_1^0 W^{(*)} b) = 0.6$. 

In Fig.~\ref{fig:excl3body_06} we see that the limits are severely weakened and much more of the stop parameter
space is unconstrained.\footnote{The current available data for the ATLAS monojet study~\cite{Aad:2014nra} do not 
allow for a reliable combination when branching ratios of less than 100\% exist in a model. In addition, 
since the study relies on charm tagging that is difficult to reliably simulate with a fast detector 
simulation we remove the study from this figure.} The $W^+W^-$ measurement is still effective however and
allows us to constrain stop masses up to $\sim 180$~GeV for a LSP mass of $\sim 80$~GeV. In fact for low stop masses
we can successfully exclude models with $m_{\tilde{\chi}^0_1}<60$~GeV which current searches are not sensitive to.

\section{Summary\label{sec:summary}}

We analysed constraints on the stop sector in light of the recent measurement 
of $W^+W^-$ production cross section by CMS. We show that this measurement provides 
constraints on light top squarks that are complementary to the dedicated
LHC searches. 

The best sensitivity is obtained along the 
line where an intermediate decay-mediating $W$ boson becomes on-shell, where the conventional
stop searches have a particular weakness. Assuming that this the only 
available decay mode, the reach is $\mstop{1} \gtrsim 230 \gev$. The method retains its 
sensitivity even for significantly reduced branching ratios to the analysed final state. We demonstrate 
that in case of $BR = 0.6$, stops with masses $\mstop{1} \lesssim 180 \gev$ are also excluded. We note 
that in the reduced branching fraction scenario, the other searches are significantly limited and this additionally
shows the complementarity of these approaches.

\section*{Note added}
After completing this study, a summary of ATLAS Run-1 stop searches has been published~\cite{Aad:2015pfx}. It 
includes a dedicated stop search along similar lines to the suggestion in this Letter. We note 
that our results for 0-jet SR are consistent with those presented in ref.~\cite{Aad:2015pfx}. However, in some parts 
of the parameter space, in particular for $\mstop{1} \simeq m_t, \mneu{1} \simeq 0$, the CMS 1-jet SR~\cite{CMS-PAS-SMP-14-016}  
offers a stronger bound.

\section*{Acknowledgements}
\noindent KR has been supported by the MINECO (Spain) under contract FPA2013-44773-P; 
Consolider-Ingenio CPAN\linebreak CSD2007-00042; the Spanish MINECO Centro de excelencia Severo Ochoa Program under grant SEV-2012-0249; and by JAE-Doc program.

%\section*{References}

\bibliographystyle{utphys}
\bibliography{stop_excl}

\providecommand{\href}[2]{#2}\begingroup\raggedright\begin{thebibliography}{10}

\bibitem{Aad:2013ija}
{\bfseries ATLAS} Collaboration, G.~Aad {\em et~al.}, ``{Search for direct
  third-generation squark pair production in final states with missing
  transverse momentum and two $b$-jets in $\sqrt{s} =$ 8 TeV $pp$ collisions
  with the ATLAS detector},''
  \href{http://dx.doi.org/10.1007/JHEP10(2013)189}{{\em JHEP} {\bfseries 1310}
  (2013) 189},
\href{http://arxiv.org/abs/1308.2631}{{\ttfamily arXiv:1308.2631 [hep-ex]}}.
%%CITATION = ARXIV:1308.2631;%%.

\bibitem{Aad:2014kra}
{\bfseries ATLAS} Collaboration, G.~Aad {\em et~al.}, ``{Search for top squark
  pair production in final states with one isolated lepton, jets, and missing
  transverse momentum in $\sqrt s =$8 TeV $pp$ collisions with the ATLAS
  detector},'' \href{http://dx.doi.org/10.1007/JHEP11(2014)118}{{\em JHEP}
  {\bfseries 1411} (2014) 118},
\href{http://arxiv.org/abs/1407.0583}{{\ttfamily arXiv:1407.0583 [hep-ex]}}.
%%CITATION = ARXIV:1407.0583;%%.

\bibitem{Aad:2014bva}
{\bfseries ATLAS} Collaboration, G.~Aad {\em et~al.}, ``{Search for direct pair
  production of the top squark in all-hadronic final states in proton-proton
  collisions at $\sqrt{s}=8$ TeV with the ATLAS detector},''
  \href{http://dx.doi.org/10.1007/JHEP09(2014)015}{{\em JHEP} {\bfseries 1409}
  (2014) 015},
\href{http://arxiv.org/abs/1406.1122}{{\ttfamily arXiv:1406.1122 [hep-ex]}}.
%%CITATION = ARXIV:1406.1122;%%.

\bibitem{Aad:2014qaa}
{\bfseries ATLAS} Collaboration, G.~Aad {\em et~al.}, ``{Search for direct
  top-squark pair production in final states with two leptons in pp collisions
  at $\sqrt{s} =$ 8 TeV with the ATLAS detector},''
  \href{http://dx.doi.org/10.1007/JHEP06(2014)124}{{\em JHEP} {\bfseries 1406}
  (2014) 124},
\href{http://arxiv.org/abs/1403.4853}{{\ttfamily arXiv:1403.4853 [hep-ex]}}.
%%CITATION = ARXIV:1403.4853;%%.

\bibitem{Aad:2014nra}
{\bfseries ATLAS} Collaboration, G.~Aad {\em et~al.}, ``{Search for
  pair-produced third-generation squarks decaying via charm quarks or in
  compressed supersymmetric scenarios in $pp$ collisions at $\sqrt{s}=8~$TeV
  with the ATLAS detector},''
  \href{http://dx.doi.org/10.1103/PhysRevD.90.052008}{{\em Phys.Rev.}
  {\bfseries D90} no.~5, (2014) 052008},
\href{http://arxiv.org/abs/1407.0608}{{\ttfamily arXiv:1407.0608 [hep-ex]}}.
%%CITATION = ARXIV:1407.0608;%%.

\bibitem{Aad:2014mfk}
{\bfseries ATLAS} Collaboration, G.~Aad {\em et~al.}, ``{Measurement of Spin
  Correlation in Top-Antitop Quark Events and Search for Top Squark Pair
  Production in pp Collisions at $\sqrt{s}=8$ TeV Using the ATLAS Detector},''
\href{http://arxiv.org/abs/1412.4742}{{\ttfamily arXiv:1412.4742 [hep-ex]}}.
%%CITATION = ARXIV:1412.4742;%%.

\bibitem{CMS-PAS-SUS-13-009}
{\bfseries CMS} Collaboration, ``{Search for top squarks decaying to a charm
  quark and a neutralino in events with a jet and missing transverse
  momentum},'' Tech. Rep. CMS-PAS-SUS-13-009, CERN, Geneva, 2014.
\newblock \url{http://cds.cern.ch/record/1644584}.

\bibitem{CMS-PAS-SUS-13-015}
{\bfseries CMS} Collaboration, ``{Search for top squarks in multijet events
  with large missing momentum in proton-proton collisions at 8 TeV},'' Tech.
  Rep. CMS-PAS-SUS-13-015, CERN, Geneva, 2013.
\newblock \url{http://cds.cern.ch/record/1635353}.

\bibitem{Chatrchyan:2013xna}
{\bfseries CMS} Collaboration, S.~Chatrchyan {\em et~al.}, ``{Search for
  top-squark pair production in the single-lepton final state in pp collisions
  at $\sqrt{s}$ = 8 TeV},''
  \href{http://dx.doi.org/10.1140/epjc/s10052-013-2677-2}{{\em Eur.Phys.J.}
  {\bfseries C73} no.~12, (2013) 2677},
\href{http://arxiv.org/abs/1308.1586}{{\ttfamily arXiv:1308.1586 [hep-ex]}}.
%%CITATION = ARXIV:1308.1586;%%.

\bibitem{Khachatryan:2015pwa}
{\bfseries CMS} Collaboration, V.~Khachatryan {\em et~al.}, ``{Search for
  supersymmetry using razor variables in events with $b$-tagged jets in $pp$
  collisions at $\sqrt{s} =$ 8 TeV},''
  \href{http://dx.doi.org/10.1103/PhysRevD.91.052018}{{\em Phys.Rev.}
  {\bfseries D91} (2015) 052018},
\href{http://arxiv.org/abs/1502.00300}{{\ttfamily arXiv:1502.00300 [hep-ex]}}.
%%CITATION = ARXIV:1502.00300;%%.

\bibitem{Khachatryan:2015wza}
{\bfseries CMS} Collaboration, V.~Khachatryan {\em et~al.}, ``{Searches for
  third generation squark production in fully hadronic final states in
  proton-proton collisions at $\sqrt{s}=8$ TeV},''
\href{http://arxiv.org/abs/1503.08037}{{\ttfamily arXiv:1503.08037 [hep-ex]}}.
%%CITATION = ARXIV:1503.08037;%%.

\bibitem{CMS-PAS-SUS-14-021}
{\bfseries CMS} Collaboration, ``{Search for supersymmetry in events with soft
  leptons, low jet multiplicity, and missing transverse momentum in
  proton-proton collisions at $\sqrt{s} = 8$ TeV},'' Tech. Rep.
  CMS-PAS-SUS-14-021, CERN, Geneva, 2015.
\newblock \url{http://cds.cern.ch/record/2010110}.

\bibitem{atlas_combined}
{\bfseries ATLAS} Collaboration.
  \url{https://atlas.web.cern.ch/Atlas/GROUPS/PHYSICS/CombinedSummaryPlots/SUSY/ATLAS_SUSY_Stop_tLSP/ATLAS_SUSY_Stop_tLSP.png}.

\bibitem{cms_combined}
{\bfseries CMS} Collaboration.
  \url{https://twiki.cern.ch/twiki/pub/CMSPublic/SUSYSMSSummaryPlots8TeV/T2tt_ICHEP2014_All.pdf}.

\bibitem{Papucci:2011wy}
M.~Papucci, J.~T. Ruderman, and A.~Weiler, ``{Natural SUSY Endures},''
  \href{http://dx.doi.org/10.1007/JHEP09(2012)035}{{\em JHEP} {\bfseries 1209}
  (2012) 035},
\href{http://arxiv.org/abs/1110.6926}{{\ttfamily arXiv:1110.6926 [hep-ph]}}.
%%CITATION = ARXIV:1110.6926;%%.

\bibitem{Czakon:2014fka}
M.~Czakon, A.~Mitov, M.~Papucci, J.~T. Ruderman, and A.~Weiler, ``{Closing the
  stop gap},'' \href{http://dx.doi.org/10.1103/PhysRevLett.113.201803}{{\em
  Phys.Rev.Lett.} {\bfseries 113} no.~20, (2014) 201803},
\href{http://arxiv.org/abs/1407.1043}{{\ttfamily arXiv:1407.1043 [hep-ph]}}.
%%CITATION = ARXIV:1407.1043;%%.

\bibitem{Eifert:2014kea}
T.~Eifert and B.~Nachman, ``{Sneaky light stop},''
  \href{http://dx.doi.org/10.1016/j.physletb.2015.02.039}{{\em Phys.Lett.}
  {\bfseries B743} (2015) 218--223},
\href{http://arxiv.org/abs/1410.7025}{{\ttfamily arXiv:1410.7025 [hep-ph]}}.
%%CITATION = ARXIV:1410.7025;%%.

\bibitem{Ferretti:2015dea}
G.~Ferretti, R.~Franceschini, C.~Petersson, and R.~Torre, ``{Spot the stop with
  a $b$-tag},''
\href{http://arxiv.org/abs/1502.01721}{{\ttfamily arXiv:1502.01721 [hep-ph]}}.
%%CITATION = ARXIV:1502.01721;%%.

\bibitem{Delgado:2012eu}
A.~Delgado, G.~F. Giudice, G.~Isidori, M.~Pierini, and A.~Strumia, ``{The light
  stop window},'' \href{http://dx.doi.org/10.1140/epjc/s10052-013-2370-5}{{\em
  Eur.Phys.J.} {\bfseries C73} no.~3, (2013) 2370},
\href{http://arxiv.org/abs/1212.6847}{{\ttfamily arXiv:1212.6847 [hep-ph]}}.
%%CITATION = ARXIV:1212.6847;%%.

\bibitem{Buckley:2014fqa}
M.~R. Buckley, T.~Plehn, and M.~J. Ramsey-Musolf, ``{Top squark with mass close
  to the top quark},'' \href{http://dx.doi.org/10.1103/PhysRevD.90.014046}{{\em
  Phys.Rev.} {\bfseries D90} no.~1, (2014) 014046},
\href{http://arxiv.org/abs/1403.2726}{{\ttfamily arXiv:1403.2726 [hep-ph]}}.
%%CITATION = ARXIV:1403.2726;%%.

\bibitem{Batell:2015zla}
B.~Batell and S.~Jung, ``{Probing Light Stops with Stoponium},''
\href{http://arxiv.org/abs/1504.01740}{{\ttfamily arXiv:1504.01740 [hep-ph]}}.
%%CITATION = ARXIV:1504.01740;%%.

\bibitem{Curtin:2012nn}
D.~Curtin, P.~Jaiswal, and P.~Meade, ``{Charginos Hiding In Plain Sight},''
  \href{http://dx.doi.org/10.1103/PhysRevD.87.031701}{{\em Phys.Rev.}
  {\bfseries D87} no.~3, (2013) 031701},
\href{http://arxiv.org/abs/1206.6888}{{\ttfamily arXiv:1206.6888 [hep-ph]}}.
%%CITATION = ARXIV:1206.6888;%%.

\bibitem{Rolbiecki:2013fia}
K.~Rolbiecki and K.~Sakurai, ``{Light stops emerging in $WW$ cross section
  measurements?},'' \href{http://dx.doi.org/10.1007/JHEP09(2013)004}{{\em JHEP}
  {\bfseries 1309} (2013) 004},
\href{http://arxiv.org/abs/1303.5696}{{\ttfamily arXiv:1303.5696 [hep-ph]}}.
%%CITATION = ARXIV:1303.5696;%%.

\bibitem{Kim:2014eva}
J.~S. Kim, K.~Rolbiecki, K.~Sakurai, and J.~Tattersall, ``{'Stop' that
  ambulance! New physics at the LHC?},''
  \href{http://dx.doi.org/10.1007/JHEP12(2014)010}{{\em JHEP} {\bfseries 1412}
  (2014) 010},
\href{http://arxiv.org/abs/1406.0858}{{\ttfamily arXiv:1406.0858 [hep-ph]}}.
%%CITATION = ARXIV:1406.0858;%%.

\bibitem{Curtin:2014zua}
D.~Curtin, P.~Meade, and P.-J. Tien, ``{Natural SUSY in Plain Sight},''
  \href{http://dx.doi.org/10.1103/PhysRevD.90.115012}{{\em Phys.Rev.}
  {\bfseries D90} no.~11, (2014) 115012},
\href{http://arxiv.org/abs/1406.0848}{{\ttfamily arXiv:1406.0848 [hep-ph]}}.
%%CITATION = ARXIV:1406.0848;%%.

\bibitem{Dermisek:2015vra}
R.~Dermisek, E.~Lunghi, and S.~Shin, ``{Contributions of flavor violating
  couplings of a Higgs boson to $pp\to WW$},''
\href{http://arxiv.org/abs/1503.08829}{{\ttfamily arXiv:1503.08829 [hep-ph]}}.
%%CITATION = ARXIV:1503.08829;%%.

\bibitem{ATLAS:2012mec}
{\bfseries ATLAS} Collaboration, G.~Aad {\em et~al.}, ``{Measurement of
  $W^+W^-$ production in pp collisions at $\sqrt{s}=7$ TeV with the ATLAS
  detector and limits on anomalous $WWZ$ and $WW\gamma$ couplings},''
  \href{http://dx.doi.org/10.1103/PhysRevD.87.112001,
  10.1103/PhysRevD.88.079906}{{\em Phys.Rev.} {\bfseries D87} no.~11, (2013)
  112001},
\href{http://arxiv.org/abs/1210.2979}{{\ttfamily arXiv:1210.2979 [hep-ex]}}.
%%CITATION = ARXIV:1210.2979;%%.

\bibitem{Chatrchyan:2013oev}
{\bfseries CMS} Collaboration, S.~Chatrchyan {\em et~al.}, ``{Measurement of
  $W^+W^-$ and $ZZ$ production cross sections in pp collisions at $\sqrt{s} =
  8$ TeV},'' \href{http://dx.doi.org/10.1016/j.physletb.2013.03.027}{{\em
  Phys.Lett.} {\bfseries B721} (2013) 190--211},
\href{http://arxiv.org/abs/1301.4698}{{\ttfamily arXiv:1301.4698 [hep-ex]}}.
%%CITATION = ARXIV:1301.4698;%%.

\bibitem{ATLAS-CONF-2014-033}
{\bfseries ATLAS} Collaboration, ``{Measurement of the $W^+W^-$ production
  cross section in proton-proton collisions at $\sqrt{s} =8$ TeV with the ATLAS
  detector},'' Tech. Rep. ATLAS-CONF-2014-033, CERN, Geneva, Jul, 2014.
\newblock \url{http://cds.cern.ch/record/1728248}.

\bibitem{Gehrmann:2014fva}
T.~Gehrmann, M.~Grazzini, S.~Kallweit, P.~Maierhöfer, A.~von Manteuffel, {\em
  et~al.}, ``{$W^+W^-$ Production at Hadron Colliders in Next to Next to
  Leading Order QCD},''
  \href{http://dx.doi.org/10.1103/PhysRevLett.113.212001}{{\em Phys.Rev.Lett.}
  {\bfseries 113} no.~21, (2014) 212001},
\href{http://arxiv.org/abs/1408.5243}{{\ttfamily arXiv:1408.5243 [hep-ph]}}.
%%CITATION = ARXIV:1408.5243;%%.

\bibitem{Jaiswal:2014yba}
P.~Jaiswal and T.~Okui, ``{Explanation of the $WW$ excess at the LHC by
  jet-veto resummation},''
  \href{http://dx.doi.org/10.1103/PhysRevD.90.073009}{{\em Phys.Rev.}
  {\bfseries D90} no.~7, (2014) 073009},
\href{http://arxiv.org/abs/1407.4537}{{\ttfamily arXiv:1407.4537 [hep-ph]}}.
%%CITATION = ARXIV:1407.4537;%%.

\bibitem{Meade:2014fca}
P.~Meade, H.~Ramani, and M.~Zeng, ``{Transverse momentum resummation effects in
  $W^+W^-$ measurements},''
  \href{http://dx.doi.org/10.1103/PhysRevD.90.114006}{{\em Phys.Rev.}
  {\bfseries D90} no.~11, (2014) 114006},
\href{http://arxiv.org/abs/1407.4481}{{\ttfamily arXiv:1407.4481 [hep-ph]}}.
%%CITATION = ARXIV:1407.4481;%%.

\bibitem{CMS-PAS-SMP-14-016}
{\bfseries CMS} Collaboration, ``{Measurement of the $W^+W^-$ cross section in
  pp collisions at $\sqrt{s} = 8$ TeV and limits on anomalous gauge couplings
  },'' Tech. Rep. CMS-PAS-SMP-14-016, CERN, Geneva, 2015.
\newblock \url{http://cds.cern.ch/record/2002016}.

\bibitem{Hikasa:1987db}
K.-i. Hikasa and M.~Kobayashi, ``{Light Scalar Top at $e^+ e^-$ Colliders},''
\href{http://dx.doi.org/10.1103/PhysRevD.36.724}{{\em Phys.Rev.} {\bfseries
  D36} (1987) 724}.
%%CITATION = PHRVA,D36,724;%%.

\bibitem{Muhlleitner:2011ww}
M.~Muhlleitner and E.~Popenda, ``{Light Stop Decay in the MSSM with Minimal
  Flavour Violation},'' \href{http://dx.doi.org/10.1007/JHEP04(2011)095}{{\em
  JHEP} {\bfseries 1104} (2011) 095},
\href{http://arxiv.org/abs/1102.5712}{{\ttfamily arXiv:1102.5712 [hep-ph]}}.
%%CITATION = ARXIV:1102.5712;%%.

\bibitem{Porod:1998yp}
W.~Porod, ``{More on higher order decays of the lighter top squark},''
  \href{http://dx.doi.org/10.1103/PhysRevD.59.095009}{{\em Phys.Rev.}
  {\bfseries D59} (1999) 095009},
\href{http://arxiv.org/abs/hep-ph/9812230}{{\ttfamily arXiv:hep-ph/9812230
  [hep-ph]}}.
%%CITATION = HEP-PH/9812230;%%.

\bibitem{Boehm:1999tr}
C.~Boehm, A.~Djouadi, and Y.~Mambrini, ``{Decays of the lightest top squark},''
  \href{http://dx.doi.org/10.1103/PhysRevD.61.095006}{{\em Phys.Rev.}
  {\bfseries D61} (2000) 095006},
\href{http://arxiv.org/abs/hep-ph/9907428}{{\ttfamily arXiv:hep-ph/9907428
  [hep-ph]}}.
%%CITATION = HEP-PH/9907428;%%.

\bibitem{Grober:2014aha}
R.~Grober, M.~Muhlleitner, E.~Popenda, and A.~Wlotzka, ``{Light Stop Decays:
  Implications for LHC Searches},''
\href{http://arxiv.org/abs/1408.4662}{{\ttfamily arXiv:1408.4662 [hep-ph]}}.
%%CITATION = ARXIV:1408.4662;%%.

\bibitem{Grober:2015fia}
R.~Grober, M.~Muhlleitner, E.~Popenda, and A.~Wlotzka, ``{Light stop decays
  into $W b \tilde{\chi}_1^0$ near the kinematic threshold},''
\href{http://arxiv.org/abs/1502.05935}{{\ttfamily arXiv:1502.05935 [hep-ph]}}.
%%CITATION = ARXIV:1502.05935;%%.

\bibitem{Rolbiecki:2009hk}
K.~Rolbiecki, J.~Tattersall, and G.~Moortgat-Pick, ``{Towards Measuring the
  Stop Mixing Angle at the LHC},''
  \href{http://dx.doi.org/10.1140/epjc/s10052-010-1517-x}{{\em Eur.Phys.J.}
  {\bfseries C71} (2011) 1517},
\href{http://arxiv.org/abs/0909.3196}{{\ttfamily arXiv:0909.3196 [hep-ph]}}.
%%CITATION = ARXIV:0909.3196;%%.

\bibitem{Alwall:2014hca}
J.~Alwall, R.~Frederix, S.~Frixione, V.~Hirschi, F.~Maltoni, {\em et~al.},
  ``{The automated computation of tree-level and next-to-leading order
  differential cross sections, and their matching to parton shower
  simulations},'' \href{http://dx.doi.org/10.1007/JHEP07(2014)079}{{\em JHEP}
  {\bfseries 1407} (2014) 079},
\href{http://arxiv.org/abs/1405.0301}{{\ttfamily arXiv:1405.0301 [hep-ph]}}.
%%CITATION = ARXIV:1405.0301;%%.

\bibitem{Sjostrand:2006za}
T.~Sjostrand, S.~Mrenna, and P.~Z. Skands, ``{PYTHIA 6.4 Physics and Manual},''
  \href{http://dx.doi.org/10.1088/1126-6708/2006/05/026}{{\em JHEP} {\bfseries
  0605} (2006) 026},
\href{http://arxiv.org/abs/hep-ph/0603175}{{\ttfamily arXiv:hep-ph/0603175
  [hep-ph]}}.
%%CITATION = HEP-PH/0603175;%%.

\bibitem{Beenakker:1997ut}
W.~Beenakker, M.~Kramer, T.~Plehn, M.~Spira, and P.~Zerwas, ``{Stop production
  at hadron colliders},''
  \href{http://dx.doi.org/10.1016/S0550-3213(98)00014-5}{{\em Nucl.Phys.}
  {\bfseries B515} (1998) 3--14},
\href{http://arxiv.org/abs/hep-ph/9710451}{{\ttfamily arXiv:hep-ph/9710451
  [hep-ph]}}.
%%CITATION = HEP-PH/9710451;%%.

\bibitem{Beenakker:2010nq}
W.~Beenakker, S.~Brensing, M.~Kramer, A.~Kulesza, E.~Laenen, {\em et~al.},
  ``{Supersymmetric top and bottom squark production at hadron colliders},''
  \href{http://dx.doi.org/10.1007/JHEP08(2010)098}{{\em JHEP} {\bfseries 1008}
  (2010) 098},
\href{http://arxiv.org/abs/1006.4771}{{\ttfamily arXiv:1006.4771 [hep-ph]}}.
%%CITATION = ARXIV:1006.4771;%%.

\bibitem{Drees:2013wra}
M.~Drees, H.~Dreiner, D.~Schmeier, J.~Tattersall, and J.~S. Kim, ``{CheckMATE:
  Confronting your Favourite New Physics Model with LHC Data},''
\href{http://arxiv.org/abs/1312.2591}{{\ttfamily arXiv:1312.2591 [hep-ph]}}.
%%CITATION = ARXIV:1312.2591;%%.

\bibitem{Kim:2015wza}
J.~S. Kim, D.~Schmeier, J.~Tattersall, and K.~Rolbiecki, ``{A framework to
  create customised LHC analyses within CheckMATE},''
\href{http://arxiv.org/abs/1503.01123}{{\ttfamily arXiv:1503.01123 [hep-ph]}}.
%%CITATION = ARXIV:1503.01123;%%.

\bibitem{deFavereau:2013fsa}
{\bfseries DELPHES 3} Collaboration, J.~de~Favereau {\em et~al.}, ``{DELPHES 3,
  A modular framework for fast simulation of a generic collider experiment},''
  \href{http://dx.doi.org/10.1007/JHEP02(2014)057}{{\em JHEP} {\bfseries 1402}
  (2014) 057},
\href{http://arxiv.org/abs/1307.6346}{{\ttfamily arXiv:1307.6346 [hep-ex]}}.
%%CITATION = ARXIV:1307.6346;%%.

\bibitem{Cacciari:2011ma}
M.~Cacciari, G.~P. Salam, and G.~Soyez, ``{FastJet User Manual},''
  \href{http://dx.doi.org/10.1140/epjc/s10052-012-1896-2}{{\em Eur.Phys.J.}
  {\bfseries C72} (2012) 1896},
\href{http://arxiv.org/abs/1111.6097}{{\ttfamily arXiv:1111.6097 [hep-ph]}}.
%%CITATION = ARXIV:1111.6097;%%.

\bibitem{Cacciari:2008gp}
M.~Cacciari, G.~P. Salam, and G.~Soyez, ``{The Anti-$k_t$ jet clustering
  algorithm},'' \href{http://dx.doi.org/10.1088/1126-6708/2008/04/063}{{\em
  JHEP} {\bfseries 0804} (2008) 063},
\href{http://arxiv.org/abs/0802.1189}{{\ttfamily arXiv:0802.1189 [hep-ph]}}.
%%CITATION = ARXIV:0802.1189;%%.

\bibitem{Bahr:2008pv}
M.~Bahr, S.~Gieseke, M.~Gigg, D.~Grellscheid, K.~Hamilton, {\em et~al.},
  ``{Herwig++ Physics and Manual},''
  \href{http://dx.doi.org/10.1140/epjc/s10052-008-0798-9}{{\em Eur.Phys.J.}
  {\bfseries C58} (2008) 639--707},
\href{http://arxiv.org/abs/0803.0883}{{\ttfamily arXiv:0803.0883 [hep-ph]}}.
%%CITATION = ARXIV:0803.0883;%%.

\bibitem{Bellm:2013lba}
J.~Bellm, S.~Gieseke, D.~Grellscheid, A.~Papaefstathiou, S.~Platzer, {\em
  et~al.}, ``{Herwig++ 2.7 Release Note},''
\href{http://arxiv.org/abs/1310.6877}{{\ttfamily arXiv:1310.6877 [hep-ph]}}.
%%CITATION = ARXIV:1310.6877;%%.

\bibitem{Nadolsky:2008zw}
P.~M. Nadolsky, H.-L. Lai, Q.-H. Cao, J.~Huston, J.~Pumplin, {\em et~al.},
  ``{Implications of CTEQ global analysis for collider observables},''
  \href{http://dx.doi.org/10.1103/PhysRevD.78.013004}{{\em Phys.Rev.}
  {\bfseries D78} (2008) 013004},
\href{http://arxiv.org/abs/0802.0007}{{\ttfamily arXiv:0802.0007 [hep-ph]}}.
%%CITATION = ARXIV:0802.0007;%%.

\bibitem{Lai:2010vv}
H.-L. Lai, M.~Guzzi, J.~Huston, Z.~Li, P.~M. Nadolsky, {\em et~al.}, ``{New
  parton distributions for collider physics},''
  \href{http://dx.doi.org/10.1103/PhysRevD.82.074024}{{\em Phys.Rev.}
  {\bfseries D82} (2010) 074024},
\href{http://arxiv.org/abs/1007.2241}{{\ttfamily arXiv:1007.2241 [hep-ph]}}.
%%CITATION = ARXIV:1007.2241;%%.

\bibitem{0954-3899-28-10-313}
A.~L. Read, ``Presentation of search results: the cl's technique,'' {\em
  J.Phys.G} {\bfseries 28} no.~10, (2002) 2693.

\bibitem{Baak:2014wma}
M.~Baak, G.~Besjes, D.~Côte, A.~Koutsman, J.~Lorenz, {\em et~al.},
  ``{HistFitter software framework for statistical data analysis},''
  \href{http://dx.doi.org/10.1140/epjc/s10052-015-3327-7}{{\em Eur.Phys.J.}
  {\bfseries C75} no.~4, (2015) 153},
\href{http://arxiv.org/abs/1410.1280}{{\ttfamily arXiv:1410.1280 [hep-ex]}}.
%%CITATION = ARXIV:1410.1280;%%.

\bibitem{Cranmer:2012sba}
{\bfseries ROOT} Collaboration, K.~Cranmer, G.~Lewis, L.~Moneta, A.~Shibata,
  and W.~Verkerke,
``{HistFactory: A tool for creating statistical models for use with RooFit and
  RooStats},''.
%%CITATION = CERN-OPEN-2012-016 ETC.;%%.

\bibitem{Moneta:2010pm}
L.~Moneta, K.~Belasco, K.~S. Cranmer, S.~Kreiss, A.~Lazzaro, {\em et~al.},
  ``{The RooStats Project},'' {\em PoS} {\bfseries ACAT2010} (2010) 057,
\href{http://arxiv.org/abs/1009.1003}{{\ttfamily arXiv:1009.1003
  [physics.data-an]}}.
%%CITATION = ARXIV:1009.1003;%%.

\bibitem{Verkerke:2003ir}
W.~Verkerke and D.~P. Kirkby, ``{The RooFit toolkit for data modeling},'' {\em
  eConf} {\bfseries C0303241} (2003) MOLT007,
\href{http://arxiv.org/abs/physics/0306116}{{\ttfamily arXiv:physics/0306116
  [physics]}}.
%%CITATION = PHYSICS/0306116;%%.

\bibitem{Brun:1997pa}
R.~Brun and F.~Rademakers, ``{ROOT: An object oriented data analysis
  framework},''
\href{http://dx.doi.org/10.1016/S0168-9002(97)00048-X}{{\em Nucl.Instrum.Meth.}
  {\bfseries A389} (1997) 81--86}.
%%CITATION = NUIMA,A389,81;%%.

\bibitem{Antcheva:2011zz}
I.~Antcheva, M.~Ballintijn, B.~Bellenot, M.~Biskup, R.~Brun, {\em et~al.},
  ``{ROOT: A C++ framework for petabyte data storage, statistical analysis and
  visualization},''
\href{http://dx.doi.org/10.1016/j.cpc.2011.02.008}{{\em Comput.Phys.Commun.}
  {\bfseries 182} (2011) 1384--1385}.
%%CITATION = CPHCB,182,1384;%%.

\bibitem{Aad:2015pfx}
{\bfseries ATLAS} Collaboration, G.~Aad {\em et~al.}, ``{ATLAS Run 1 searches
  for direct pair production of third-generation squarks at the Large Hadron
  Collider},''
\href{http://arxiv.org/abs/1506.08616}{{\ttfamily arXiv:1506.08616 [hep-ex]}}.
%%CITATION = ARXIV:1506.08616;%%.

\end{thebibliography}\endgroup

\end{document}